\documentclass[a4paper,11pt]{article}
\usepackage{jinstpub} % for details on the use of the package, please see the JINST-author-manual
\usepackage{lineno}
\usepackage{colortbl}

\title{\boldmath Design and \textcolor{red}{D}evelopment of the P-cubed Target Insertion Device (P$^3$-TID) }

\author[a,1]{R. Mena-Andrade,}
\author[a]{J-L. Grenard,}
\author[a]{K. Guergar,}
\author[a]{R. Seidenbinder,}
\author[b]{M.I. Besana,}
\author[a]{\\\textcolor{red}{B. Humann,}}
\author[a]{\textcolor{red}{G. Lavezzari,}}
\author[a]{\textcolor{red}{M. Widorski,}}
\author[a]{\textcolor{red}{A. Lechner,}}
\author[a,b]{N. Vallis,}
\author[b]{M. Zykova,}
\author[b]{\\D. Hauenstein,}
\author[b]{R. Zennaro,}
\author[b]{P. Craievich,}
\author[a,1]{and A. Perillo-Marcone\note{Corresponding author.}}
\affiliation[a]{European Organization for Nuclear Research (CERN),\\
Geneva, Switzerland}
\affiliation[b]{Paul Scherrer Institute (PSI),\\
Villigen, Switzerland}

\emailAdd{ramiro.francisco.mena.andrade@cern.ch, antonio.perillo-marcone@cern.ch}

\abstract{The P-cubed Target Insertion Device (P$^3$-TID) is a research instrument  dedicated to test novel positron source target configurations inside of the proof-of-principle PSI Positron Production (P-cubed or P$^3$) experiment at the Paul Scherrer Institute. The device allows an easy installation, positioning and replacement of different fixed targets. The present article describes its mechanical design at a detailed level.}


\arxivnumber{2605.19650} % Only if you have one
\begin{document}
\maketitle
\flushbottom

\section{Introduction}
\label{sec:01_introduction}

\begin{figure}[ht!]
\includegraphics[width=\textwidth]{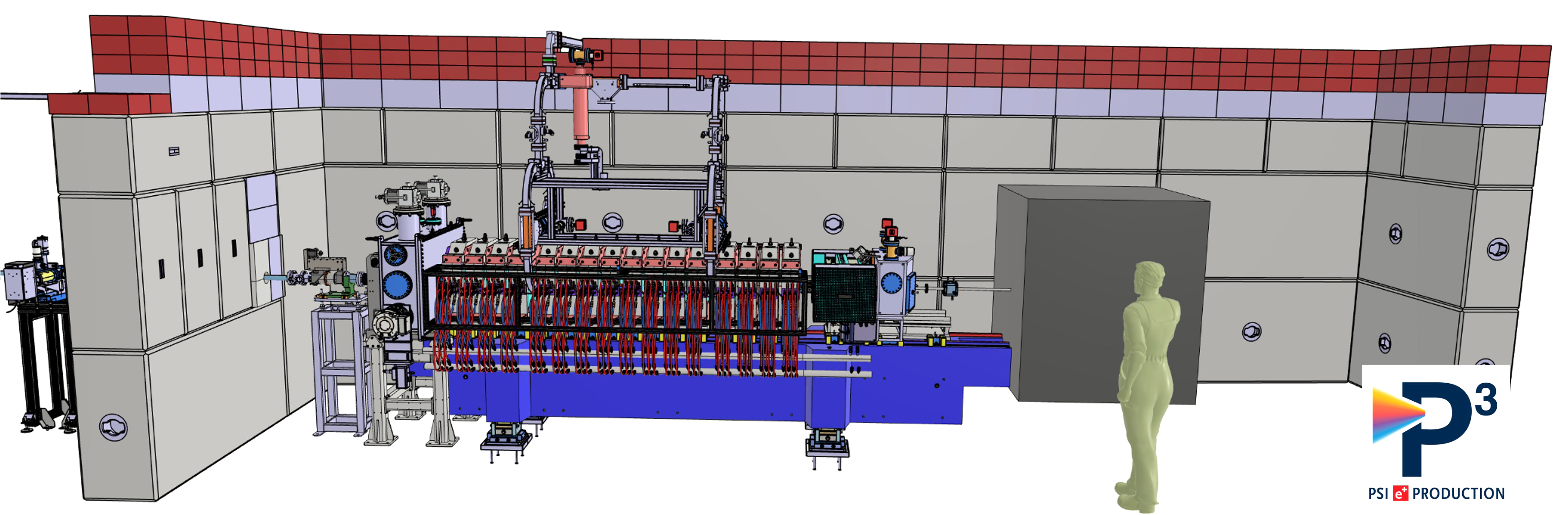}
\caption{Representation of the PSI Positron Production (P$^3$) Experiment inside of the SwissFEL facility \cite{Benedikt2025}\label{fig:P3_layout}}
\end{figure}

%The P$^3$ experiment at PSI}
The Paul Scherrer Institut (PSI) Positron Production experiment (referred as P$^3$ or P-cubed) is a proof-of-principle positron source to be used as a demonstrator for the Future Circular Collider (FCC-ee) in its first stage as a lepton collider \cite{Benedikt2025}. It will be hosted in the SwissFEL facility where a primary electron (e$^-$) beam with an energy up to 6 GeV will be used to hit a fixed target made of a high-Z material to produce positrons (e$^+$) using the pair-production mechanism \cite{Hubbell2006,CHART2023}. A representation of the experiment layout is shown in Figure \ref{fig:P3_layout}.

\begin{figure}[ht!]
\includegraphics[width=\textwidth]{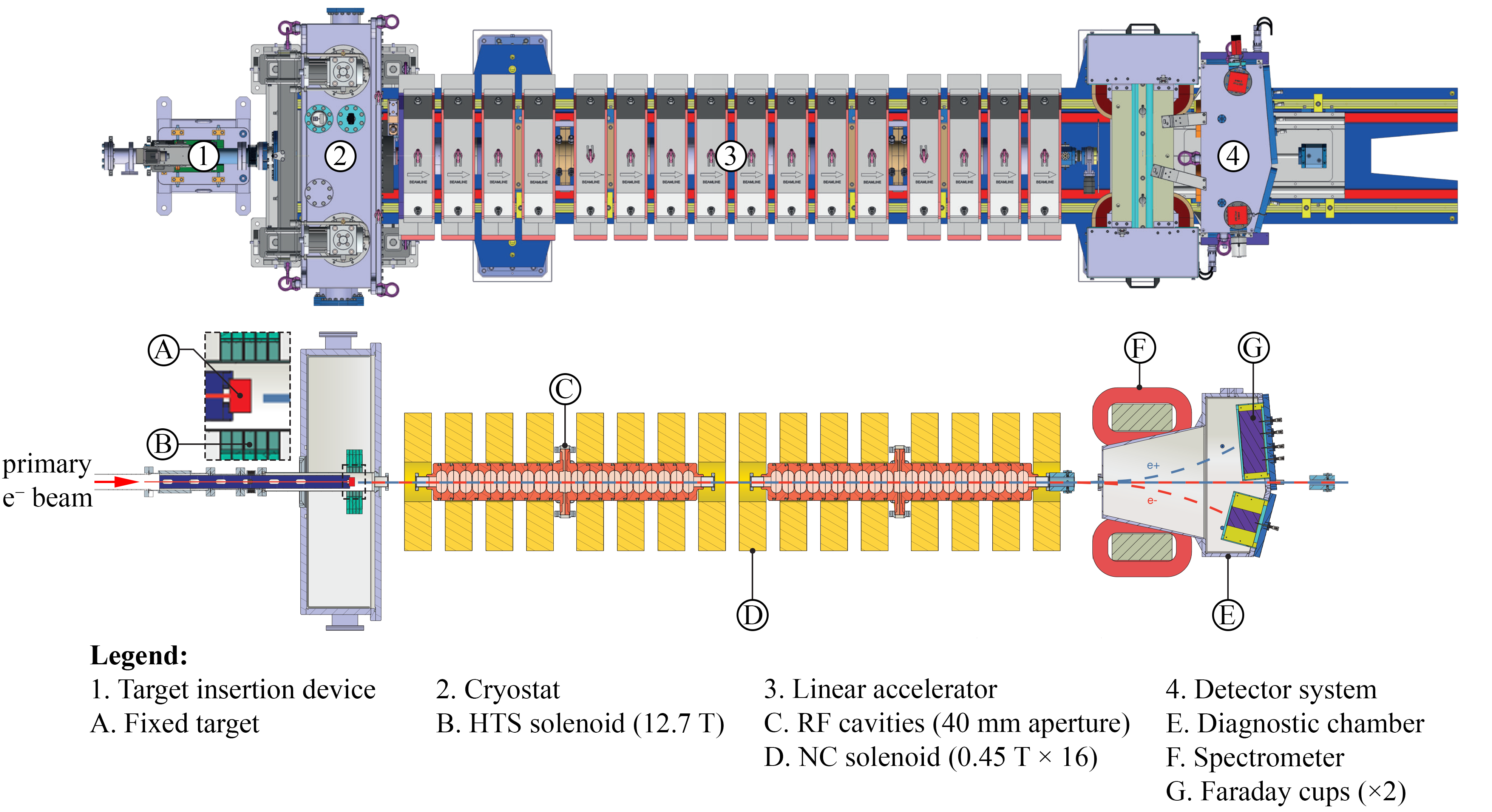}
\caption{The PSI Positron Production (P$^3$) Experiment. The top view includes the subsystems (top) while the cross section shows the main components of each subsystem (bottom). Adapted from \cite{Vallis2024b}\label{fig:P3_components}}
\end{figure}

%P3 summary
At a high-level, the main subsystems of the P$^3$ experiment can be summarized as: a positron source, contained in the target insertion device (1), a capture system installed inside of a cryostat (2), the linac (3) and the detector system (4), as shown in Figure \ref{fig:P3_components} (top).

%Key technologies
The novel capture system to be tested in P$^3$ is based in two key technologies: \emph{i)} a high-temperature superconducting (HTS) solenoid and \emph{ii)} the use of two standing-wave (SW) large aperture Radio Frequency (RF) cavities in S-band. While the high magnetic field (12.7 T) provided by the HTS around the target region will capture the primary positrons, the use of the RF cavities with a 40 mm diameter of iris aperture is devoted to catch the secondary positrons with the assistance of 16 normal conducting solenoids with a constant magnetic field (0.45 T) along the beam axis. Integrating both technologies maximizes capture efficiency, enabling a tenfold increase in positron yield over state-of-the-art facilities like SuperKEKB, which currently relies on a conventional, normal-conducting Flux Concentrator (FC) \cite{Enomoto2021}. For a detailed overview of the P$^3$ experiment, with emphasis on the capture system and the beam dynamics, the interested reader is referred to \cite{Vallis2024}.

%Context
The P$^3$ Target Insertion Device (P$^3$-TID) is a research instrument designed and developed at CERN in collaboration with PSI, under the CHART (Swiss Accelerator Research and Technology) collaboration program. The resulting device allows an easy installation, positioning and replacement of different fixed targets with the aim to test novel positron source target configurations inside of the  P$^3$ experiment.

%Document structure
This paper is devoted to present the mechanical design of the P$^3$-TID, complementing the information provided on the P$^3$ experiment in the article of \mbox{N. Vallis \emph{et al.}, \cite{Vallis2024}}. The document is organized as follows: section \ref{sec:02_target_design} is focused to the instrument design, first explaining the similitudes and differences in terms of beam parameters between P$^3$ and FCC-ee, then an overview of the main components is provided.  Then, section \ref{sec:03_Radioprotection} covers the radioprotection aspects of the instrument inside of the P$^3$ experiment setup, followed by a description of the target exchange procedure. Next, the foreseen future upgrades for the P$^3$-TID are introduced in section \ref{sec:04_Upgrades}. Finally, the conclusions and directions for future work are highlighted in section  \ref{sec:05_conclusions}.

\section{P$^3$-TID design}
\label{sec:02_target_design}

\subsection{FCC-ee and P$^3$: beam parameters comparison}
\label{ssec:beam_parameters}

%Intro
The design of a target as a beam intercepting device is mainly driven by the incoming beam properties. For comparison purposes, Table \ref{tab:i} contains a summary of the primary e$^-$ beam parameters for FCC-ee and P$^3$. Here, it can be seen that P$^3$ will be operated under less demanding beam conditions, using a scaling factor $\eta$ such that $\frac{1}{15200} \leq \eta \leq\frac{3}{36608}$, relative to the FCC-ee configuration. This design criteria is mainly translated into a lower beam power available with the aim to cope with the stringent radiation protection limits at SwissFEL. As a result, the P$^3$ target is subjected to minor thermo-mechanical loads and radiation damage.

%Motivation from R&D point of view
From the beam parameters comparison, it can be concluded that due to the low average power deposited on the target, the presence of a dedicated cooling system for the target is not mandatory. However, the P$^3$ experiment presents a unique opportunity to carry out manufacturing R$\&$D activities to be implemented in FCC-ee.

\subsection{Design overview}
\label{ssec:design_overview}

\begin{figure}[!ht]
\centering
\includegraphics[width=0.625\textwidth]{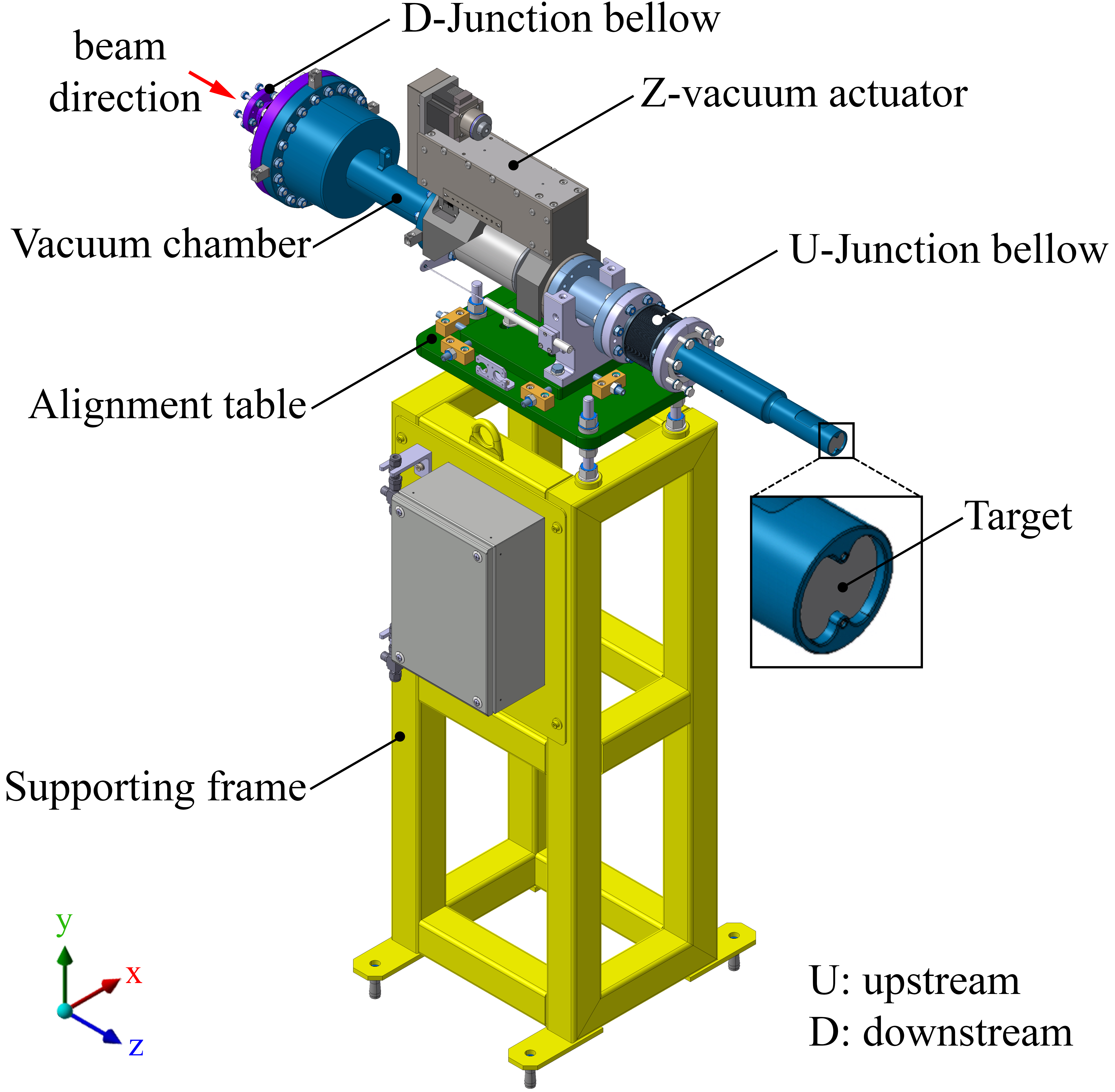}
\caption{P$^3$ target inserting device (P$^3$-TID) overview.\label{fig:i}}
\end{figure}

Following the design specifications \cite{Mena2024}, the resulting P$^3$-TID allows an easy installation, positioning and replacement of different fixed targets. Figure \ref{fig:i} shows the CAD model of the instrument where the main subsystems/components are pointed out. In the assembly, the fixed target is the core of the device. This part is located inside of a vacuum chamber that longitudinally traverse the instrument. Located in the middle, the z-vacuum actuator provides movement to the assembly which rests on an adjustable table, which in turn is mounted on its corresponding supporting frame. More details are provided below.

\begin{table}[!hb]
\centering
\caption{Summary of the e$^-$ beam parameters for FCC-ee and P$^3$ \label{tab:i}}
\smallskip
\begin{tabular}{lcc}
\hline
&\textbf{FCC-ee}&\textbf{P$^3$}\\
\hline
Injection energy E$_0$ [GeV]& 2.86 & 2.86~-~6\\
Beam size rms $\sigma_x$ [mm] & 1 & 0.5~-~1\\
Bunch charge [nC] &3.8 & 0.2\\
Repetition rate [Hz] & 100 & 1\\
Bunches per pulse [-] &4&1\\
Beam power [W] & 4347.2 & 0.572~-~1.20\\
Average power on target [W] &1000& 0.148~-~0.31\\
\hline
\end{tabular}
\end{table}

\newpage

\subsection{Target types}

The instrument is capable to accept different types of positron targets. A total of six different target types were manufactured. The main goal is to test the impact of changing geometry, material and primary e$^-$ beam in the e$^+$ production. A brief description for each possible configuration is discussed bellow.

\paragraph{Material type:} As the e$^+$ production depends on the material cross-section, high-Z materials are preferred for the positron source target. Using tungsten (Z=74) as reference, a target of tantalum (Z=73) was manufactured to compare its response. Both beam intercepting devices have a cylindrical geometry. As a remark, gold (Z=79) was considered during the design phase \cite{Mena2024b}, but due to the low thermo-mechanical properties in combination with its high material price, this option was not implemented for P$^3$.

\paragraph{Primary e$^-$ beam energy:} This is not a problem for the SwissFEL linac, so, 2.86 GeV and 6 GeV cases will be tested with the corresponding targets made with the radiation length $X_0$ of 4.5 and 5 for each considered candidate material (W and Ta), respectively.
\paragraph{Geometry:} Using the cylindrical tungsten target as a baseline, two conical targets without cooling were manufactured to quantify the e$^+$ yield improvement based on geometry and beam size terms. A detailed analysis in terms of the conical geometry optimization with respect to the e$^+$ yield  and its thermo-mechanical response under the FCC-ee primary e$^-$ beam parameters is described in  \cite{Vallis&Mena2025}. Although it was found that the conical $\sigma_x=0.5$~mm target can not withstand the FCC-ee conditions at its current design version, the resulting geometry is still of interest for P$^3$.

\begin{figure}
    \centering
    \includegraphics[width=\textwidth]{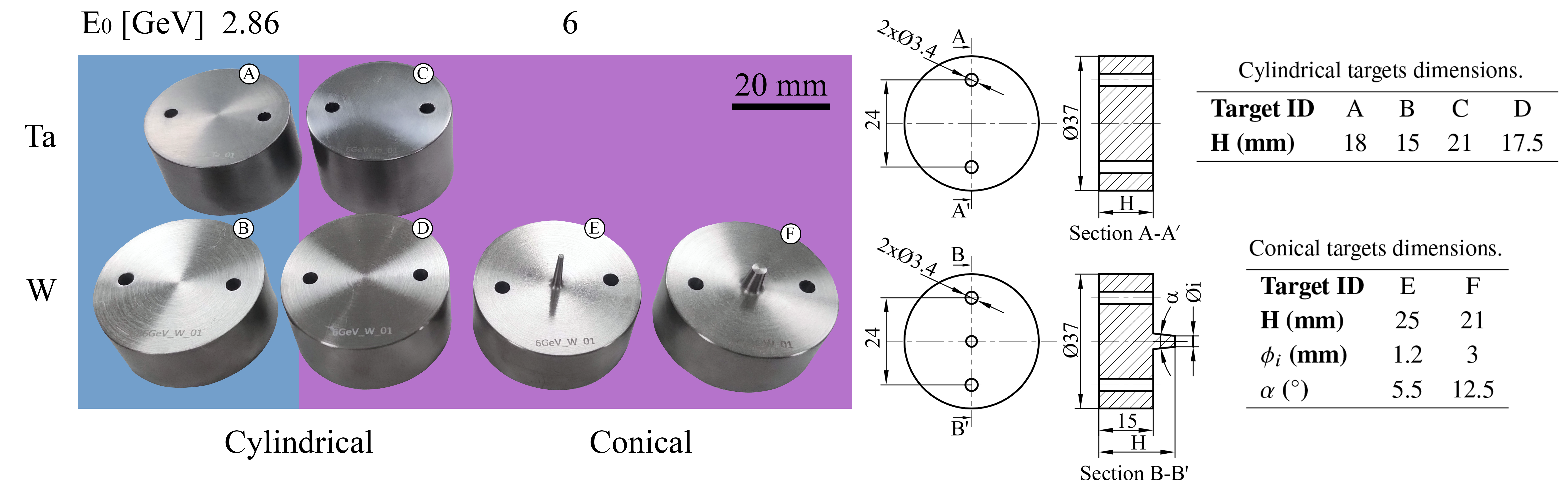}
    \caption{Available targets to be tested at P$^3$: (left) manufactured units and (right) main dimensions.}\label{fig:Target_types}
\end{figure}

%Summary and possible extensions
Figure \ref{fig:Target_types} (left) provides a graphical overview of the different target configurations to be tested during the experimental campaign at P$^3$. Each target is assigned a unique identifier from A to F, and a spare unit is available for each device. Furthermore, the modular design allows for the incorporation of new target geometries for future upgrades (see section \ref{sec:04_Upgrades}), provided that the outer diameter of $\phi$37~mm is maintained. The main dimensions of the different target geometries are summarized in Figure \ref{fig:Target_types} (right). 

\subsection{The actuation system}

\begin{figure}[!ht]
    \centering
    \includegraphics[width=\linewidth]{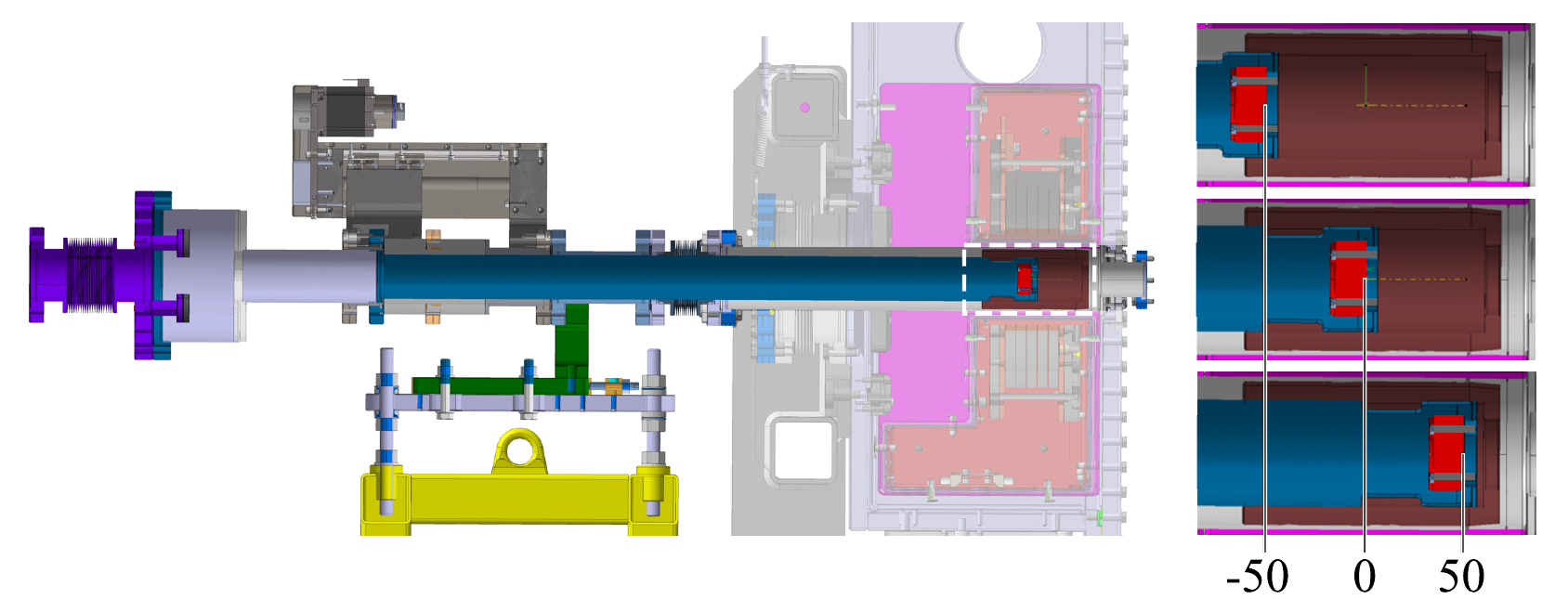}
    \caption{Cross section of the P$^3$-TID installed in the cryostat (left). The detailed zones show the $\pm$ 50 mm stroke and the corresponding positions of the target (right).}
    \label{fig:Target_stroke}
\end{figure}

The instrument is capable to move along the primary e$^-$ beam axis (z-axis) thanks to the use of a vacuum compatible z-actuator \cite{PV2026}. This mechatronic component has a flexible body made of a bellow and it is used in combination to a set of two junction bellows, referred as downstream and upstream in Figure \ref{fig:P3_components}, allowing a motion range of $\pm$50~mm with a precision of 0.1~mm.\footnote{The design is inspired in the collimators used in the Large Hadron Collider (LHC) at CERN that uses a similar solution to allow the movement of its jaws and absorb any possible misalignment during installation and operation \cite{Bertarelli2004}.} The goal of this feature is to study the influence of the target position inside of the HTS solenoid and scan the location where the e$^+$ yield production is maximized. Figure \ref{fig:Target_stroke} shows the cross section of the instrument and the different positions of the target. In addition, the design includes a limit switch sensor acting on both sides to protect the actuator in case of malfunction (e.g stroke above the allowed movement range). 

%Fatigue remark
From the operational point of view, the actuator system must ensure a minimum of 100 cycles, in other words, its use is not extensive as once a target position is chosen, its current location will be associated to a specific measurement test. In addition, as both the z-actuator and the bellow are certified to provide a service life of $10^4$ cycles \cite{PV2026},  fatigue is not a main concern for the assembly.

\begin{figure}[!ht]
    \centering
    \includegraphics[width=\linewidth]{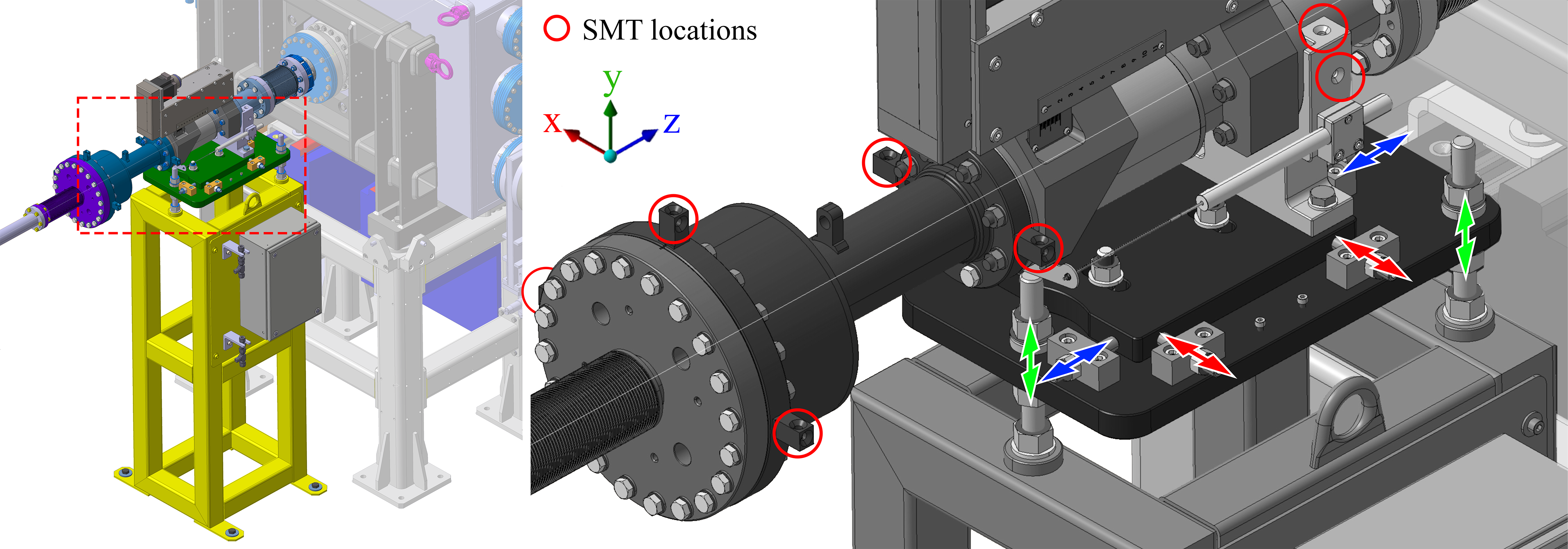}
    \caption{Alignment table and positioning system:  overview (left), location of the slots to install the survey measurement targets (SMT) and the available degrees of freedom (DOF’s) of the alignment table (right). Note: each color of the arrows represents the corresponding DOF’s with respect to the coordinate system.}
    \label{fig:P3-position}
\end{figure}

\subsection{Vacuum chamber}
The vacuum chamber has a cylindrical geometry with one stepped section, just after the downstream junction bellow (see Figure \ref{fig:Target_stroke}(left)) that allows the installation of the P$^3$-TID subsystem to the Porthos switchyard’s beam pipe.\footnote{This geometry is meant to provide the required space to allow the insertion and extraction of two cooling \textcolor{red}{tubes} for a future upgrade. See Section \ref{sec:04_Upgrades} and Figure \ref{fig:Target_with_cooling} (left).} Following the e$^-$ beam direction, the part is inserted into the z-acutator, going through the upstream junction bellow until it is properly placed inside of the cryostat. Then, an intermediate flange serves as an interface to connect the part to the inlet region of the z-actuator. Finally, the part host the fixed target at is end, as shown in Figure \ref{fig:Target_stroke} (right). 

\subsection{Alignment table and positioning system}
Once the P$^3$-TID is connected to the beam pipe and the target is located inside of the cryostat, it is required to calibrate its position with respect to the rest of the P$^3$ devices. This step is of vital importance so that the fixed target is in the right location inside of the experiment. To this end, the adjustment table counts with 3 independent screws systems to move the assembly along the x, y and z axis. In addition, to allow the triangulation of the device, multiple slots are available in the vacuum chamber ($\times$3), z-vacuum actuator ($\times$2) and the U-support frame ($\times$2) to install the survey measurement targets (SMT). Both features are shown in Figure \ref{fig:P3-position}.

\section{Radioprotection aspects}
\label{sec:03_Radioprotection}
\subsection{Overview}
\label{ssec:3.1_Overview}

Radiation Protection (RP) assessments for the P$^3$ experiment at PSI were conducted using FLUKA simulations \cite{FLUKA2026, FLUKA2015, FLUKA2022, FLUKA2024, FLUKA2024b}. The performed studies modeled only the 6 GeV electron beam case with a bunch charge of 200 pC at a repetition rate of 1 Hz, impacting target configurations C-E. Two irradiation scenarios were considered. In the first case, the target was irradiated during a single machine shift, and exposed to beam particles over a continuous period of 30 hours. For the second case, the same target was used for an entire year, undergoing irradiation during six machine shifts spaced 60 days apart, with no beam in between. During each shift, the target was exposed to beam particles for 30 hours \cite{Besana2026}.  A summary of the different RP studies and the maximum residual dose rate after a cooling period of 30 days is presented in Table \ref{tab:ii}.

\begin{figure}[!ht]
    \centering
    \includegraphics[width=\linewidth]{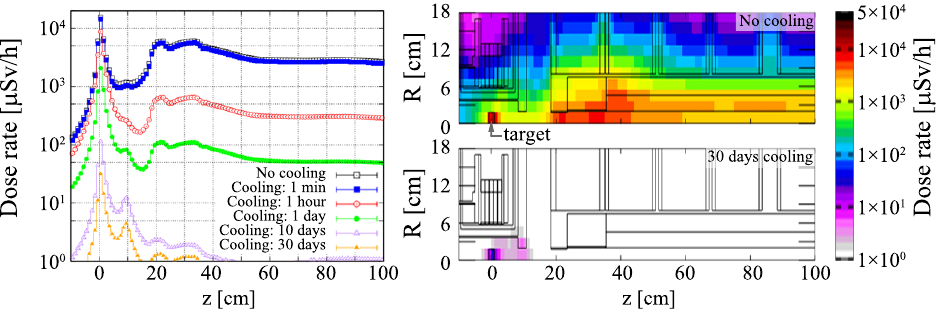}
    \caption{Residual dose rate maps in the P$^3$ experimental complex using target D under the irradiation period of 6 $\times$ 30 hours in $\mu$Sv/h. Longitudinal profile after irradiation period (no cooling time) and for multiple cooling times (left). (Top) Radial distribution  at the end of the irradiation period  (no cooling time) and (bottom) after 30 days of cooling time (right). Adapted from \cite{Besana2026}.}
    \label{fig:Dose_rate}
\end{figure}

From the 4 cases analyzed, the tungsten cylindrical configuration (D) under the subjected 6$\times$30 hours irradiation scenario presented the highest residual dose. As a result, the target is the hot spot in the entire experimental complex, with a peak dose rate of 1.57$\times$10$^4$ $\mu$Sv/h, as depicted in the longitudinal profile of Figure \ref{fig:Dose_rate} (left). Then, multiple cooling scenarios were considered and after one month of cooling time, the maximum residual dose rate at the target is estimated at 35 $\mu$Sv/h. This level allows for manual maintenance under Controlled Zone protocols. 

\begin{table}[ht!]
\centering
\caption{Maximum \textcolor{red}{simulated} residual dose rate values after a cooling time of 30 days \label{tab:ii}}
\smallskip
\begin{tabular}{cccccc}
\hline
\textbf{Target} & \textbf{Material} & \textbf{Geometry} &  \textbf{$\sigma_x$} &\textbf{Irradiation scenario}  &\textbf{Max. residual} \\
\textbf{ID}     &                   &   &\textbf{[mm]}&\textbf{[shifts~$\times$~hours]}       &\textbf{dose rate [$\mu$Sv/h]}\\
\hline
C & Ta  & cylindrical &1&1~$\times$~30& 33\\
D & W   & cylindrical &1&1~$\times$~30 & 16\\
E & W   & conical     &0.5&1~$\times$~30 & 31\\
D & W   & cylindrical &1&6~$\times$~30 & 35\\
\hline
\end{tabular}
\end{table}

\begin{table}[ht!]
\centering
\caption{Steps for removal (replacement) of a target \label{tab:iii}}
\smallskip
\begin{tabular}{cl}
\hline
\textbf{Step}&\textbf{Task}\\
\hline
a & Open (close) the vacuum chamber downstream bellow\\
b & Open (close) the Z-actuator flange\\
c & Remove (install) a portion of the beam pipe in the downstream section\\
d & Extract (insert) the vacuum chamber\\
e & Remove (install) the two screw\textcolor{red}{s} that support the target\\
f & Remove (install) the target\\
\hline
\end{tabular}
\end{table}

\begin{figure}[ht!]
\centering
\includegraphics[width=.47\textwidth]{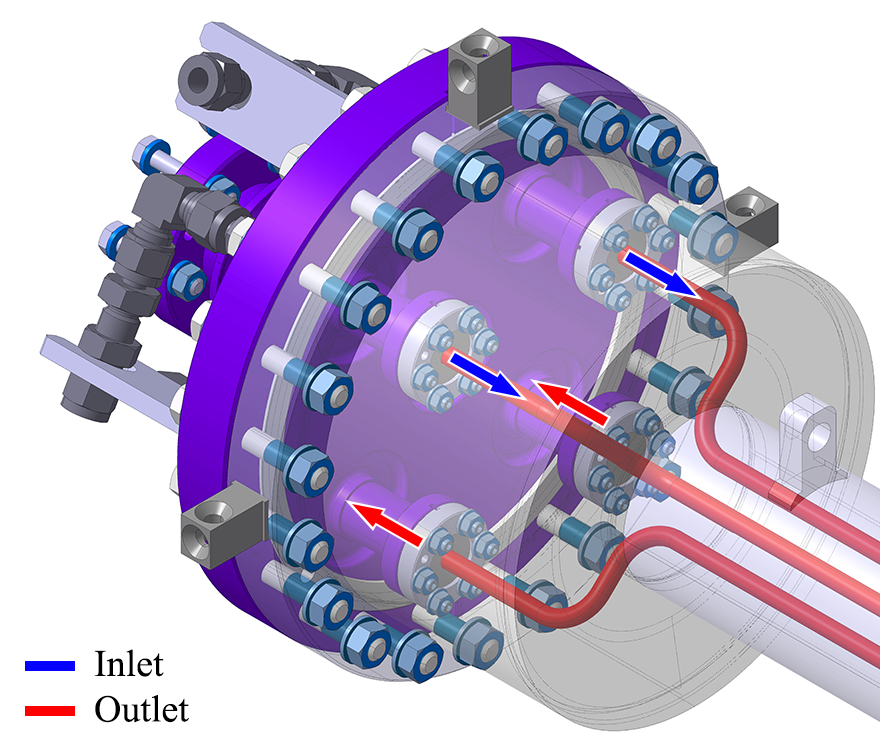}
\qquad
\includegraphics[width=.47\textwidth]{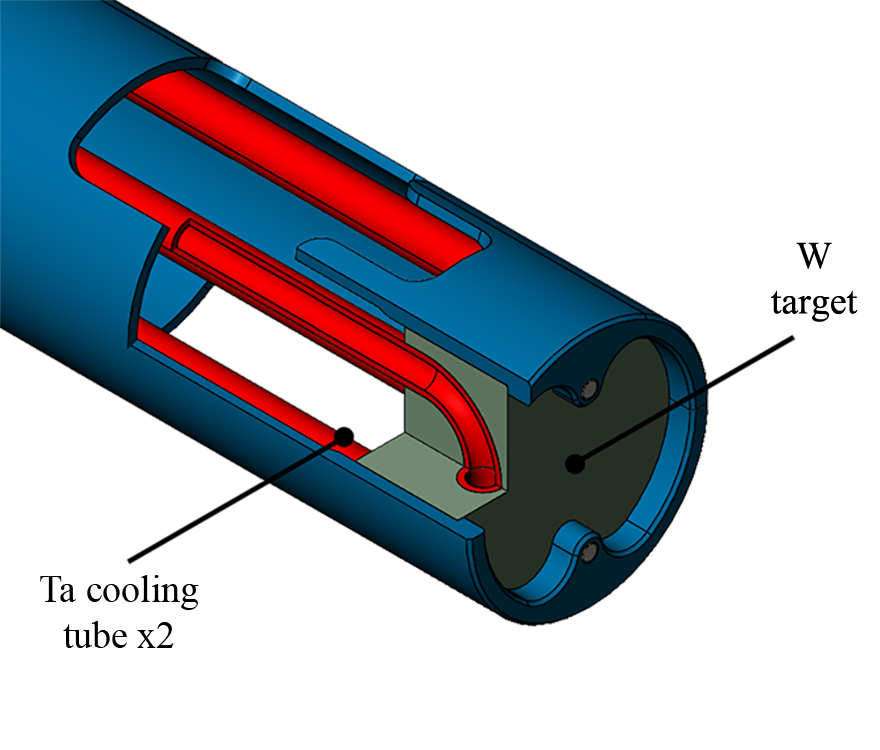}
\caption{Tungsten target with cooling system: (left) detail at the vacuum chamber and (right) cross-section.\label{fig:Target_with_cooling}}
\end{figure}

\subsection{Target exchange}
\textcolor{red}{As defined in Section \ref{ssec:3.1_Overview}, a 30-day cool-down period with the beam off is required before accessing the P$^3$ area for target exchange. During this time, the cryostat remains under a continuous maintenance regime.  The HTS does not need to be warmed up; however, the solenoid current must be turned off to eliminate the hazard of working in a high magnetic field. Additionally, because the beam pipe and cryostat vacuum systems are completely isolated from one another, venting the cryostat is unnecessary.  Finally, the target exchange is performed according to the sequence outlined in Table \ref{tab:iii}, while installing a new target uses the reverse procedure.} 

%The steps involved are given in Table  The installation of a new target follows the same procedure but in reverse.

\section{Future planned upgrades}
\label{sec:04_Upgrades}

\paragraph{Target with cooling:} A target unit with embedded tantalum cooling \textcolor{red}{tubes} is under development. While the expected thermal load coming from the SwissFEL electron gun is not a main concern, as discussed in Section \ref{ssec:beam_parameters}, the goal of this prototype is to test the manufacturing R$\&$D for FCC-ee. \textcolor{red}{Tantalum was selected for the target cooling system due to its excellent mechanical and corrosion resistance properties, as well as its proven diffusion bonding compatibility with tungsten via Hot Isostatic Pressing (HIP) \cite{Busom2020}.} Figure \ref{fig:Target_with_cooling} (left) shows the details of the tantalum cooling \textcolor{red}{tubes} inlet and outlet located at the vacuum chamber level while Figure \ref{fig:Target_with_cooling} (right)  presents a cross section of the target installed at the other end. A partial cross section is done for representation purposes to show the location of the embedded elbow that is placed as close as possible to the exit surface of the target. This feature is done to properly dissipate the heat under the FCC-ee conditions. A detailed study was carried out to define the manufacturing parameters of the cooling system \cite{Mena2026}, and a first version of the \textcolor{red}{HIP} capsule was fabricated. However, up to date the proper machining of the grooves in the tungsten core to host the variable geometry of the tantalum cooling \textcolor{red}{tubes}' elbow is the current bottleneck. A future publication will address the challenges an lessons learned in detail.

\section{Conclusions}
\label{sec:05_conclusions}

%Summary
The P$^3$ experiment is a proof-of-concept positron source and capture system with the potential to improve the current state-of-the-art positron yield. In this paper, the P$^3$-TID was introduced as the subsystem responsible for integrating the positron source target across all other P$^3$ system components. Its mechanical design was conceived to ensure flexibility when carrying out tests for different fixed target configurations, including the effect of the target’s position within the HTS solenoid capture system, and its impact on the high field RF cavities. These two technologies are essential for enhancing the positron yield production.

%Future work
The P$^3$-TID was successfully installed in autumn 2025 at PSI. Future work will include the beam commissioning, experiment setup and run of dedicated test campaigns. A future planned upgrade is under development to validate the manufacturing route for the positron source target prototype for FCC-ee.
\acknowledgments
%acknowlegdements
This work was done under the auspices of CHART (Swiss Accelerator Research and Technology) and the Future Circular Collider Innovation Study (FCCIS). This project has received funding from the European Union’s Horizon 2020 research and innovation programme under grant agreement No 951754.

% Bibliography

%% [A] Recommended: using JHEP.bst file
\bibliographystyle{JHEP}
\bibliography{biblio.bib}

@inproceedings{Enomoto2021,
  author       = {Enomoto, Yoshinori and Abe, Keiko and Okada, Naoki and Takatomi, Toshikazu},
  title        = {{A New Flux Concentrator Made of Cu Alloy for the SuperKEKB Positron Source}},
  booktitle    = {\href{https://jacow.org/ipac2021/papers/wepab144.pdf}{Proc. IPAC'21}},
  pages        = {2954-2956},
  eid          = {WEPAB144},
  year         = {2021},
}

@article{Vallis2024,
  title = {Proof-of-principle ${e}^{+}$ source for future colliders},
  author = {Vallis, N. and Craievich, P. and Sch\"ar, M. and Zennaro, R. and Auchmann, B. and Braun, H. H. and Besana, M. I. and Duda, M. and Fortunati, R. and Garcia-Rodrigues, H. and Hauenstein, D. and Ischebeck, R. and Ismaili, E. and Jurani\ifmmode \acute{c}\else \'{c}\fi{}, P. and Kosse, J. and Magazinik, A. and Marcellini, F. and Michlmayr, T. and M\"uller, S. and Pedrozzi, M. and Rotundo, R. and Orlandi, G. L. and Seidel, M. and Strohmaier, N. and Zykova, M.},
  journal = {Phys. Rev. Accel. Beams},
  volume = {27},
  issue = {1},
  pages = {013401},
  numpages = {16},
  year = {2024},
  month = {Jan},
  publisher = {American Physical Society},
  doi = {10.1103/PhysRevAccelBeams.27.013401},
  url = {https://link.aps.org/doi/10.1103/PhysRevAccelBeams.27.013401}
}

@techreport{Mena2024,
    author = {Mena-Andrade, R. and Grenard, J-L.},
    title = {Design specifications for the {P$^3$} target at {PSI}},
    institution = {CERN EDMS 2873767},
    year = 2024,
    note= {EDMS 2873767}
}

@article{Vallis&Mena2025,
    title = {Conical {T}argets for {E}nhanced {H}igh-{C}urrent {P}ositron {S}ources},
    author = {Vallis, N. and Mena-Andrade, R. and Humann, B. and Zhao, Y. and Craievich, P. and Grenard, J-L. and Latina, A. and Lechner, A. and Perillo-Marcone, A. and Zennaro, R. and Alharthic, F. and Chaikovska, I. and Chehab, R. and Seidel, M.},
    journal = {	Nucl. Instrum. Methods Phys. Res., B},
    volume={568},
    month={Novembe },
    pages={165854},
    year = {2025},
    doi={10.1016/j.nimb.2025.165854}
}

@inproceedings{Bertarelli2004,
    author = {Bertarelli, A. and Aberle, O. and Assmann, R. and Chiaveri, E. and Kurtyka, T. and Mayer, M. and Perret, R. and Sievers, P.},
    title = {{The Mechanical Design for the {LHC} Collimators}},
    booktitle = {\href{https://accelconf.web.cern.ch/e04/PAPERS/MOPLT008.PDF}{Proc. EPAC 2004}},
    year = 2004,
}

@article{Mena2026,
    title = {Manufacturability studies for the {FCC}-ee positron source target: determination of the minimum bending radius and ovalization in tantalum cooling \textcolor{red}{tube} elbows},
    author = {Mena-Andrade, R. and Crouvizier, M. and Rigaud, J-P and Coiffet, T. and Perillo-Marcone, A.},
    journal =  {Manuscript submitted for publication},
    year={2026}
}

@phdthesis{Vallis2024b, 
    author={Vallis, Nicolas}, 
    title={Proof-of-Principle Positron Source for Future Colliders}, 
    school={EPFL}, 
    address={Lausanne}, 
    year={2024}, 
    doi={\href{http://doi.org/10.5075/epfl-thesis-11349}{10.5075/epfl-thesis-11349}}, 
    url={https://infoscience.epfl.ch/handle/20.500.14299/241518}, language={en} }

@article{Benedikt2025,
    author={Benedikt, M. and Zimmermann, F. and Auchmann, B. and others},
    title={{Future Circular Collider Feasibility Study Report: Volume 2 Accelerators, technical infrastructure and safety}},
    journal={Eur. Phys. J. Spec. Top.},
    vol={234}, 
    pages={5713–6197},
    year={2025},
    doi={https://doi.org/10.1140/epjs/s11734-025-01967-4}
}

@manual{PV2026,
    author = {Pfeiffer-Vacuum}, 
    title ={{Z}-axis precision manipulator, 100 mm
stroke, motorized, stainless steel, {DN} 63 {CF}},
    year={\href{https://vacuum-shop.com/2075800/downloads/datasheets/Datasheet_420MZA063-100-m_en.pdf}{Datasheet}, acces\textcolor{red}{s}ed: 2026-04-06},
    }

@article{FLUKA2015,
    author = {Battistoni, G. and Boehlen, T. and Cerutti, F. and Chin, P. W. and Esposito, L. S. and Fass{\`o}, A. and Ferrari, A. and Lechner, A. and Empl, A. and Mairani, A. and Mereghetti, A. and Garcia Ortega, P. and Ranft, J. and Roesler, S. and Sala, P. R. and Vlachoudis, V. and Smirnov, G.},
    title={Overview of the {FLUKA} code},
    journal={Annals of Nuclear Energy},
    volume = {82},
    pages = {10\textendash18},
    year={2015},
    doi={10.1016/j.anucene.2014.11.007}}

@article{FLUKA2022,
    author = {Ahdida, C. and Bozzato, D. and Calzolari, D. and Cerutti, F. and Charitonidis, N. and Cimmino, A. and Coronetti, A. and D'Alessandro, G. L. and Donadon Servelle, A. and Esposito, L. S. and Froeschl, R. and Garc{\'i}a Al{\'i}a, R. and Gerbershagen, A. and Gilardoni, S. and Horv{\'a}th, D. and Hugo, G. and Infantino, A. and Kouskoura, V. and Lechner, A. and Lefebvre, B. and Lerner, G. and Magistris, M. and Manousos, A. and Moryc, G. and Ogallar Ruiz, F. and Pozzi, F. and Prelipcean, D. and Roesler, S. and Rossi, R. and Sabat{\'e} Gilarte, M. and Salvat Pujol, F. and Schoofs, P. and Str{\'a}nsk{\'y}, V. and Theis, C. and Tsinganis, A. and Versaci, R. and Vlachoudis, V. and Waets, A. and Widorski, M.},
    title = {New {C}apabilities of the {FLUKA} {M}ulti-{P}urpose {C}ode},
    journal = {Frontiers in Physics },
    volume=9, 
    pages=788253,
    year = {2022},
    doi={https://doi.org/10.3389/fphy.2021.788253}
}

@article{FLUKA2024,

    author = {Hugo, G. and Ahdida, C. and Bozzato, D. and Calzolari, D. and Cerutti, F. and Ciccotelli, A. and Cimmino, A. and Devienne, A. and Donadon Servelle, A. and Dyrcz, P. K. and Esposito, L. S. and Formento, A. and Froeschl, R. and Garc{\'i}a Al{\'i}a, R. and Gilardoni, S. and Gomes, A. and Horv{\'a}th, D. and Humann, B. and Infantino, A. and Lechner, A. and Lefebvre, B. and Lerner, G. and Lorenzon, T. and Lucsanyi, D. and Magistris, M. and Marin, S. and Mazzola, G. and Niang, S. and Nowak, E. and Ogallar Ruiz, F. and Potoine, J.-B. and Pozzi, F. and Prelipcean, D. and Rodin, V. and Roesler, S. and Sabat{\'e} Gilarte, M. and Sacristan Barbero, M. and Salvat Pujol, F. and Schoofs, P. and {\c{S}}erban, A.-G. and Sharankov, I. and Theis, C. and Tisi, M. and Tsinganis, A. and Versaci, R. and Vlachoudis, V. and Waets, A. and Widorski, M. and Zymak, I.},
    title = {Latest {FLUKA} developments},
    journal = {EPJ Nuclear Sciences \& Technologies},
    volume={10},
    pages={20},
    year = {2024},
    doi={10.1051/epjn/2024023}
}

@article{FLUKA2024b,
    author = {Donadon, A. and Hugo, G. and Theis, C. and Vlachoudis, V.},
    title = {{FLAIR3 – recasting simulation experiences with the Advanced Interface for FLUKA and other Monte Carlo codes}},
    journal = {EPJ Web Conf.},
    volume=302, 
    pages={11005},
    year={2024}, 
    doi={10.1051/epjconf/202430211005 }
}

@manual{FLUKA2026,
    author = {{FLUKA website}},
    year = {\href{https://fluka.cern}{https://fluka.cern}, acces\textcolor{red}{s}ed: 2026-05-17} 
}

@inproceedings{Besana2026,
    author = {Besana, M. I. and Snuverink, J. and Vallis, N. and Zennaro, R. and Craievich, P.},
    title = {{Shielding calculations and activation studies for the PSI Positron Production experiment}},
    booktitle = {\href{https://doi.org/10.18429/JACoW-IPAC2026-TUP3081}{Proc. IPAC'26}},
    year = {2026}
}

@article{Hubbell2006,
	Author = {J.H. Hubbell},
	Journal = {Radiation Physics and Chemistry},
	volume = {75},
    pages = {614-623},
	Title = {Electron–positron pair production by photons: {A} historical overview},
	Year = {2006},
    doi={10.1016/j.radphyschem.2005.10.008}
    }

@misc{CHART2023,
    author = {Craievich, P. and Bettoni, S. and Sch{\"a}r, M. and Vallis, N. and Zennaro, R. and Auchmann, B. and Braun, H. H. and Besana, M. I. and Duda, M. and Fortunati, R. and Garcia-Rodrigues, H. and Hauenstein, D. and Ischebeck, R. and Ismaili, E. and Jurani{\'c}, P. and Kosse, J. and Marcellini, F. and Michlmayr, T. and M{\"u}ller, S. and Pedrozzi, M. and Rotundo, R. and Orlandi, G. L. and Seidel, M. and Strohmaier, N. and Zykova, M. and Grudiev, A. and Latina, A. and Doebert, S. and Vostrel, Z. and Zhao, Y. and Humann, B. and Lechner, A. and Kurtulus, A. and Mena Andrade, R. and Grenard, J.-L. and Perillo-Marcone, A. and Calviani, M. and Bartmann, W. and Duthil, Y. and Bartosik, H. and Oide, K. and Zimmermann, F. and Benedikt, M. and Chaikovska, I. and Alharthi, F. and Mytrochenko, V. and Chehab, R. and Milardi, C. and De Santis, A. and Etisken, O. and Spampinati, S. and Raubenheimer, T. and Enomoto, Y.},
    title = {{FCC-ee Injector Study and the P$^3$ Project at PSI}},
    year = {\href{https://chart.ch/wp-content/uploads/2024/02/Report-2023-FCCee-Injector.pdf}{CHART Scientific Report 2023}, 2024}
}

@conference{Mena2024b,
    author = {Mena-Andrade, R. and Humann, B. and Grenard, J.-L. and Lechner, A. and Perillo-Marcone, A. and Calviani, M.},
    booktitle ={FCC Week 2024},
    title = {\href{https://indico.cern.ch/event/1298458/contributions/5977877/attachments/2874467/5035071/20240611_FCC-Week_2024_Positron_Source_Target.pdf}{Development of {FCC}-ee and {P3} positron source targets at {CERN}}},
    year = {2024}
}

@article{Busom2020,
    author = {Busom Descarrega, Josep and Calviani, Marco and Hutsch, Thomas and López Sola, Edmundo and Pérez Fontenla, Ana Teresa and Perillo Marcone, Antonio and Sgobba, Stefano and Weißgärber, Thomas},
    title = {\textcolor{red}{Application of {H}ot {I}sostatic {P}ressing ({HIP}) technology to diffusion bond refractory metals for proton beam targets and absorbers at {CERN}}},
    journal = {Material Design \& Processing Communications},
    volume = {2},
    number = {1},
    pages = {e101},
    doi = {10.1002/mdp2.101},
    year = {2020}
}

%% or
%% [B] Manual formatting (see below)
%% (i) We suggest to always provide author, title and journal data or doi:
%% in short all the informations that clearly identify a document.
%% (ii) please avoid comments such as "For a review'', "For some examples",
%% "and references therein" or move them in the text. In general, please leave only references in the bibliography and move all
%% accessory text in footnotes.
%% (iii) Also, please have only one work for each \bibitem.

%\begin{thebibliography}{99}

%\bibitem{a}
%Author,
%\emph{Title},
%\emph{J. Abbrev.} {\bf vol} (year) pg.

%\bibitem{b}
%Author,
%\emph{Title},
%arxiv:1234.5678.

%\bibitem{c}
%Author,
%\emph{Title},
%Publisher (year).

%\end{thebibliography}
\end{document}